# Silicon-doped $\beta$-$Ga_2O_3$ films grown at 1 µm/h by suboxide molecular-beam epitaxy

Kathy Azizie,[1] Felix V. E. Hensling,[1] Cameron A. Gorsak,[1] Yunjo Kim,[2]
Daniel M. Dryden,[3] M. K. Indika Senevirathna,[4] Selena Coye,[4]
Shun-Li Shang,[5] Jacob Steele[1], Patrick Vogt[1], Nicholas A. Parker,[1] Yorick A. Birkhölzer,[1]
Jonathan P. McCandless,[6] Debdeep Jena,[1,6,7] Huili G. Xing,[1,6,7] Zi-Kui Liu,[5]
Michael D. Williams,[4] Andrew J. Green,[3] Kelson Chabak,[3] Adam T. Neal,[2] Shin Mou,[2]
Michael O. Thompson,[1] Hari P. Nair,[1] and Darrell G. Schlom[1,7,8,a)]

[1] Department of Materials Science and Engineering, Cornell University, Ithaca, New York 14853, USA

[2] Air Force Research Laboratory, Materials and Manufacturing Directorate, Wright Patterson Air Force Base, Ohio 45433, USA

[3] Air Force Research Laboratory, Sensors Directorate, Wright Patterson Air Force Base, Ohio 45433, USA

[4] Department of Physics, Clark Atlanta University, Atlanta, Georgia 30314, USA

[5] Department of Material Science and Engineering, Pennsylvania State University, University Park, Pennsylvania 16802, USA

[6] School of Electrical and Computer Engineering, Cornell University, Ithaca, New York, 14853, USA

[7] Kavli Institute at Cornell for Nanoscale Science, Ithaca, New York 14853, USA

[8] Leibniz-Institut für Kristallzüchtung, Max-Born-Str. 2, 12489 Berlin, Germany

[a)] Author to whom correspondence should be addressed. Electronic mail: schlom@cornell.edu.

## Abstract

We report the use of suboxide molecular-beam epitaxy (*S*-MBE) to grow $\beta$-$Ga_2O_3$ at a growth rate of ~1 µm/h with control of the silicon doping concentration from $5\times10^{16}$ to $10^{19}$ $cm^{-3}$. In *S*-MBE, pre-oxidized gallium in the form of a molecular beam that is 99.98% $Ga_2O$, i.e., gallium suboxide, is supplied. Directly supplying $Ga_2O$ to the growth surface bypasses the rate-limiting first step of the two-step reaction mechanism involved in the growth of $\beta$-$Ga_2O_3$ by conventional MBE. As a result, a growth rate of ~1 µm/h is readily achieved at a relatively low growth temperature ($T_{sub} \approx 525$ °C), resulting in films with high structural perfection and smooth surfaces (rms roughness of < 2 nm on ~1 µm thick films). Silicon-containing oxide sources (SiO and $SiO_2$) producing an SiO suboxide molecular beam are used to dope the $\beta$-$Ga_2O_3$ layers. Temperature-dependent Hall effect measurements on a 1 µm thick film with a mobile carrier concentration of $2.7\times10^{17}$ $cm^{-3}$ reveal a room-temperature mobility of 124 $cm^2$ $V^{-1}$ $s^{-1}$ that increases to 627 $cm^2$ $V^{-1}$ $s^{-1}$ at 76 K; the silicon dopants are found to exhibit an activation energy of 27 meV. We also demonstrate working MESFETs made from these silicon-doped $\beta$-$Ga_2O_3$ films grown by *S*-MBE at growth rates of ~1 µm/h.

## Introduction

With its very high bandgap, dopability, good mobility for electrons, and the availability of large-diameter native substrate, $\beta$-$Ga_2O_3$ has emerged as a promising material for high-power electronics.[1,2] Although molecular-beam epitaxy (MBE) is the leading technique for the growth of most compound semiconductors, for the growth of $\beta$-$Ga_2O_3$ it has serious limitations and metalorganic chemical vapor deposition (MOCVD) is currently the method of choice.[3,4] A direct comparison of the best electrical properties reported for the growth of $\beta$-$Ga_2O_3$ films by MBE[5,6] and MOCVD[3] at growth rates near current limits for each are shown in Table I, revealing the shortcomings of MBE.

Conventional MBE growth of $\beta$-$Ga_2O_3$ starts with metallic gallium, which undergoes a two-step reaction to form solid $Ga_2O_3$:[7-9]

$$2\ Ga + \text{oxidant} \rightarrow Ga_2O + \text{additional oxidant} \rightarrow Ga_2O_3, \tag{1}$$

where the oxidant is either oxygen plasma or ozone. The suboxide produced by the first step, $Ga_2O$, is volatile at typical growth temperatures and, if there is insufficient oxidant present to oxidize it, it will desorb from the substrate surface diminishing the throughput of the second step. The first step is rate-limiting and even operating at the oxidant pressure limit of that hot filaments in an MBE can withstand, i.e., ~$10^{-5}$ Torr, the growth rate of $\beta$-$Ga_2O_3$ is relatively slow, in the ~0.1 µm/h range.[2,5,6] It is possible to enhance this meager growth rate by catalytic means[10] using a process referred to as metal-oxide catalyzed epitaxy (MOCATAXY)[11] or metal-exchange catalyzed molecular beam epitaxy (MEXCAT-MBE).[12] The metallic catalyst can be added indium, which can be incorporated into $\beta$-$(Al_xGa_{1-x})_2O_3$ films at the ~1% level,[11] or through the addition of a dopant, e.g., tin.[13] Another way to increase the growth rate is to bypass the first step of the two-step reaction mechanism and directly supply an incident molecular beam of $Ga_2O$, i.e., suboxide MBE (*S*-MBE),[14] rather than by supplying a molecular beam of gallium.

Using *S*-MBE increases the growth rate by overcoming the kinetic limits imposed by the first reaction shown in Eq. (1), leading to an order of magnitude higher growth rate in the ~1 µm/h range.[14] There are a couple of options to produce a molecular beam of $Ga_2O$ suboxide. One is to use solid $Ga_2O_3$ itself[15,16] and another is to use a mixture of solid $Ga_2O_3$ and liquid gallium.[17] Vapor pressure calculations show that the dominant species in the molecular beam for both methods is the suboxide $Ga_2O$.[18,19] An advantage, however, of using the $Ga_2O_3$ + gallium mixture is the far lower temperature needed to produce the $Ga_2O$ molecular beam.[18,19] This allows more types of crucibles to be used to contain the mixture (e.g., $Al_2O_3$ and BeO crucibles), increases the $Ga_2O$ flux that can be achieved (and with it the growth rate of the $\beta$-$Ga_2O_3$ film), and decreases contaminants coming from the hot crucible and effusion cell. For example, when solid $Ga_2O_3$ has been used as a source, even at source temperatures in the 1700 °C range, the maximum growth rate that has been achieved is 0.14 µm/h and the films have been contaminated with $\sim 5\times 10^{18}$ iridium atoms $cm^{-3}$ coming from the iridium crucible used to contain the very hot $Ga_2O_3$ source.[20-22] In contrast, films of $\beta$-$Ga_2O_3$ produced using *S*-MBE using a $Ga_2O_3$ + gallium mixture result in growth rates exceeding 1 µm/h, excellent crystallinity, and smooth surfaces all at a relatively low growth temperature, $T_{sub}$.[14] The disadvantage of *S*-MBE is that generating a $Ga_2O$ molecular beam at low temperature utilizes a source containing a mixture of $Ga_2O_3$ *powder* plus gallium metal.[14,17] Unfortunately, the highest purity of $Ga_2O_3$ powder that we are able to obtain commercially only has 99.999% (5N) purity, whereas 99.99999% (7N) pure gallium is the norm for conventional MBE. This lower purity raises questions about whether *S*-MBE can grow device-quality films. The direct way to answer this question is to make devices on $\beta$-$Ga_2O_3$ layers grown by *S*-MBE and see how they perform.

Studying the mobility of $\beta$-$Ga_2O_3$ films and fabricating $\beta$-$Ga_2O_3$-based devices necessitate controlled doping. Silicon is the preferred dopant for $\beta$-$Ga_2O_3$ films as it not only yields films with the highest mobility,[2,3,23-25] but has been demonstrated to provide abrupt and controlled doping over the $10^{16}$–$10^{20}$ $cm^{-3}$ range in $\beta$-$Ga_2O_3$ films grown by MOCVD.[2,3,24-27] Silicon segregates less to the surface during MBE growth than does tin,[6] making it a superior dopant and it has the advantage over germanium of a far lower diffusion coefficient in $\beta$-$Ga_2O_3$ at temperatures required for device processing.[28] The traditional MBE approach to dope with silicon is to create a silicon molecular beam by heating silicon to high temperature in an MBE source. Unfortunately, in the high oxidant pressure (~$10^{-5}$ Torr) used for the growth of $\beta$-$Ga_2O_3$ by MBE, the surface of the silicon doping source may also oxidize.[29] Whether the silicon surface gets covered over by a layer of $SiO_2$ or is able to desorb gaseous SiO at the same rate that the silicon source is oxidized and thus remain uncoated, depends on the temperature of the silicon and the oxidant pressure. For the case that the oxidant pressure is fixed at $1\times10^{-5}$ Torr, at high temperature ($T_{\mathrm{Si}} \gtrsim 830$ °C) the silicon surface is able to desorb all of the oxide that forms and remains clean, whereas at low temperature ($T_{\mathrm{Si}} \lesssim 830$ °C) the silicon surface gets coated by solid $SiO_2$.[30-32] These changes with oxidation of the silicon source result in huge changes in the flux of silicon or SiO emanating from the silicon source, thwarting controlled and reproducible doping at device-relevant concentrations.[4,5,29]

This paper tackles the challenges of growing silicon-doped $\beta$-$Ga_2O_3$ thin films by taking advantage of the suboxide forms of both silicon and gallium. In this work, we demonstrate both controllable and reproducible silicon doping of $\beta$-$Ga_2O_3$ at device-relevant concentrations, i.e., the $5\times10^{16}$ to $10^{19}$ $cm^{-3}$ range, with a growth rate of ~1 μm/h by *S*-MBE. Having overcome the doping challenge, we then evaluate the mobility of the silicon-doped $\beta$-$Ga_2O_3$ layers and demonstrate working devices made by *S*-MBE at a growth rate of ~1 μm/h.

**Experimental**

All films were grown using a Veeco GEN10 MBE system equipped with retractable and differentially pumped effusion cells that can be exchanged without venting the entire MBE system. This facilitates refilling the $Ga_2O_3$ + gallium source. Film growth took place on (0001)-oriented sapphire substrates (Kyocera) or (010)-oriented iron-doped $\beta$-$Ga_2O_3$ single-crystal substrates (Novel Crystal Technology) with 10 mm × 10 mm × 0.5 mm dimensions held in substrate holders made of Haynes® 214® alloy. The backside of each substrate was coated with a ~200 nm thick platinum layer (on top of a ~20 nm thick titanium adhesion layer) to enable it to be radiatively heated by a SiC heating element to $T_{sub}$ = 525 °C or 550 °C as measured by an optical pyrometer operating at a wavelength of 980 nm.

A detailed description of the thermodynamics and kinetics of the growth of $Ga_2O_3$ by suboxide MBE is given in prior publications.[14,33] Thermodynamic calculations in the present work were performed using the Scientific Group Thermodata Europe (SGTE) substance database (SSUB5)[34] within the Thermo-Calc software.[35]

To obtain a molecular beam of gallium suboxide, we shake the combination of $Ga_2O_3$ powder (Alfa Aesar, 99.999% purity) and molten metallic gallium (Alfa Aesar, 99.99999% purity) to produce a mixture with a molar fraction of oxygen of $x$(O) = 0.4. The $Ga_2O_3$ + gallium charge is contained within a beryllium oxide (BeO) crucible (Materion, 99.7% purity) in the MBE effusion cell. Our thermodynamic calculations estimate such a mixture heated to the 750-1000 °C range of temperatures used in this study will produce a molecular beam that is 99.98% $Ga_2O$.[14] Films were grown at a background pressure of distilled ozone (~80% ozone with the balance being $O_2$) of $P_{O_3}$ = (1-5)×$10^{-6}$ Torr. The growth rate was determined by measuring the film thickness of $\beta$-$Ga_2O_3$ calibration films grown on sapphire substrates by X-ray reflectivity (XRR). For two of the homoepitaxial $\beta$-$Ga_2O_3$ films (Samples k and l), the growth rate was corroborated by depth measurements made using a stylus profilometer on secondary-ion mass spectrometry (SIMS) craters.

To dope silicon into $\beta$-$Ga_2O_3$ films, we tried two silicon-containing oxide sources: SiO (Alfa Aesar, 99.99% purity) and $SiO_2$ (Kurt J. Lesker, 99.99% purity) contained within $Al_2O_3$ crucibles (McDanel, 99.8% purity). From vapor pressure calculations, both sources are expected to produce a molecular beam of the suboxide SiO.[19] This approach is an attempt to circumvent the challenges associated with going back and forth between the active and passive oxidation regimes of elemental silicon as the silicon source temperature is changed, as unfortunately occurs in the growth of $\beta$-$Ga_2O_3$ by conventional MBE using a silicon source.[5,29] Both SiO and $SiO_2$ have been previously used as doping sources for silicon in oxide MBE: SiO to dope $\beta$-$Ga_2O_3$ films[36,37] and $SiO_2$ to dope $\alpha$-$Al_2O_3$ films.[38]

The $SiO_x$ flux emanating from the SiO and $SiO_2$ doping sources was measured in two ways. At high SiO and $SiO_2$ source temperatures, where the $SiO_x$ flux was sufficient to build up a multi-nanometer thick film in a reasonable time, its thickness was determined by XRR. XRR measurements were made on amorphous $SiO_x$ films deposited on unheated (0001) $Al_2O_3$ or (100) MgO substrates using the calibration scheme described by Ref. 36. The silicon fluxes calculated from these XRR measurements were independent of the oxidant employed for background pressures of distilled ozone ranging from none (vacuum) up to at least $2.5\times10^{-6}$ Torr. At lower SiO and $SiO_2$ source temperatures, the $SiO_x$ flux was inferred from Hall effect measurements of silicon-doped $\beta$-$Ga_2O_3$ films. Hall effect measurements were made using a van der Pauw geometry[39] with four ohmic contacts on 1 µm thick silicon-doped $\beta$-$Ga_2O_3$ films grown by *S*-MBE at a growth rate of ~1 µm/h on iron-doped $\beta$-$Ga_2O_3$ (010) substrates that were 10 mm × 10 mm × 0.5 mm in size. For highly silicon-doped samples (greater than $1\times10^{19}$ $cm^{-3}$), contacts to the films were made by soldering indium at the corners. In order to make ohmic contact to more lightly silicon-doped films, ~90 nm of $n^+$ epitaxial $\beta$-$Ga_2O_3$ was regrown on the four corners of the films using a shadow mask in an Agnitron Agilis 100 MOCVD system. The donor density ($N_d$) in these $n^+$ layers was greater than $1\times10^{19}$ $cm^{-3}$ to ensure that the contacts are ohmic. The MOCVD reactor pressure was 50 Torr, and the substrate temperature was 630 °C. Triethylgallium (TEGa), oxygen (99.999%), and silane (25 ppm $SiH_4$ in argon) were used as precursors, with argon (99.999%) as the carrier gas. Then indium was soldered on top of the regrown $n^+$ $\beta$-$Ga_2O_3$ contacts. A Nanometrics HL5500 Hall system was used on these samples to determine the sheet carrier concentration of mobile charges at room temperature. The silicon flux was estimated under the assumption that all silicon are activated at room temperature in the $\beta$-$Ga_2O_3$ film. The temperature-dependent transport properties of selected samples (Samples a, b, and d) were measured as a function of temperature using a Lakeshore Hall system in a vacuum cryostat at an applied magnetic field of 0.5 T.

X-ray diffraction (XRD) and reflectivity (XRR) measurements were performed in a coplanar symmetric geometry using a PANalytical Empyrean system equipped with a copper anode and a hybrid incident-beam monochromator to provide Cu $K_{\alpha1}$ radiation. Rocking curves were collected in a triple-axis configuration using a 220 Ge analyzer crystal. An Asylum

Research Cypher ES atomic force microscope (AFM) was used to measure the surface roughness.

SIMS measurements were made either using a Hiden Analytical SIMS Plus Workstation or by EAG Laboratories. The Hiden Analytical SIMS Plus Workstation used an $O_2^+$ ion source as the primary beam to profile the sample. The system features a MAXIM quadrupole mass analyzer to detect and analyze the emitted secondary ions at 30° to the probe axis. The $O_2^+$ primary ion beam was oriented at 45° relative to the sample surface. The primary beam voltage and current for the analysis were 2 kV and 140 nA, respectively. The crater area, scan density, electronic gate, and oxygen flooding pressure were 400 µm × 400 µm, 100 pixels × 100 pixels, 5% of the raster area, and $4.0\times10^{-6}$ Torr, respectively. The SIMS system base pressure was ~$5.0\times10^{-10}$ Torr.

Metal-semiconductor field-effect transistors (MESFETs) were fabricated using annealed Ti/Al/Ni/Au ohmic contacts, a Ni/Au gate, and mesa isolation via reactive-ion etching,[40] with a channel length of 3 µm and a gate length of 1 µm.

**Results**

Utilizing a $Ga_2O_3$ + gallium two-phase mixture to produce a molecular beam of the suboxide $Ga_2O$, we first map out the growth rate of $\beta$-$Ga_2O_3$ on (0001) $Al_2O_3$ substrates as a function of suboxide flux under conditions yielding epitaxial films. In Fig. 1 we demonstrate that the growth rate of $\beta$-$Ga_2O_3$ is a function of substrate temperature, distilled ozone pressure, and $Ga_2O$ flux. For a constant substrate temperature and background pressure of distilled ozone, the growth rate at first increases linearly with the $Ga_2O$ flux and then plateaus. This behavior is seen in Fig. 1(a). The linear region corresponds to oxidant-rich conditions, where all of the $Ga_2O$ supplied in the incident flux is oxidized to $Ga_2O_3$ during its residence time on the substrate surface. When the $Ga_2O$ flux increases further—into the $Ga_2O$-rich regime—the excess $Ga_2O$, beyond what can be oxidized by the background pressure of distilled ozone, is desorbed from the surface due to the volatility of $Ga_2O$ at the substrate temperatures employed. This results in a plateau in the growth rate with increasing gallium suboxide $Ga_2O$ flux. The growth rate in the plateau regime depends solely on the background pressure of distilled ozone (and substrate temperature); it is an adsorption-controlled growth regime, where the insufficient ozone flux limits the adsorption of the excess $Ga_2O$ flux. As the background ozone pressure is increased, more $Ga_2O$ can be oxidized during its residence time and the growth rate is found to increase. This occurs by an extension of the oxidant-rich (linear) regime to a new plateau of the growth rate at a higher $Ga_2O$ flux, beyond which there is insufficient ozone present to oxidize all of the $Ga_2O$ to $Ga_2O_3$ during its residence time on the substrate surface. At a substrate temperature of 525 °C and background distilled ozone pressure of $P_{O_3} = 5\times10^{-6}$ Torr, we achieve growth rates as high as 2.5 μm/h on (0001) $Al_2O_3$. Similarly, when keeping the ozone pressure constant and varying the substrate temperature, the growth rate is seen (Fig. 1(b)) to decrease with increasing substrate temperature due to the decreased residence time of the $Ga_2O$ on the substrate surface as the substrate temperature is increased. The oxidant-rich and adsorption-controlled growth regimes are still observed, just shifted due to the change in residence time. These results agree with a kinetic model developed for *S*-MBE.[14,33]

Having established conditions for the epitaxial growth of $\beta$-$Ga_2O_3$ films at a rate of ~1 μm/h, we next consider doping them with silicon. For this purpose, we tried both of the oxides of silicon, first SiO and then $SiO_2$, as source materials to dope $\beta$-$Ga_2O_3$. To calibrate the flux of the SiO molecular beam, XRR was used to determine the growth rate of amorphous $SiO_x$ films. As XRR provides both a measure of the film thickness and the film density,[41] XRR spectra of smooth $SiO_x$ films enable the silicon flux to be calculated as a function of the temperature of the SiO source, $T_{SiO}$. From the XRR spectra, the density of the amorphous $SiO_x$ films deposited by the SiO source was about 1.8 g/cm$^3$, comparable to that reported in prior studies of vacuum-deposited $SiO_x$ films deposited under similar conditions.[36,42] To evaluate the stability of the SiO source and the associated reproducibility of growing silicon-doped $\beta$-$Ga_2O_3$ film over a range of desired conditions, we grew amorphous $SiO_x$ calibration films at oxidant pressures ranging from vacuum to $P_{O_3} = 5\times10^{-6}$ Torr. While the SiO source behaved well at the high values of $T_{SiO}$ used for the growth of the XRR samples, at the lower values of $T_{SiO}$ relevant to doping $\beta$-$Ga_2O_3$ films with silicon concentrations in the $10^{17}$–$10^{19}$ cm$^{-3}$ range, SIMS measurements revealed a major challenge with using SiO as a source. At lower $T_{SiO}$, the flux is not reproducible. Not only did doping at the same value of $T_{SiO}$ vary considerably from growth to growth (as seen in Hall measurements), but often the doping concentrations measured in SIMS stacks would not follow an Arrhenius relationship as $T_{SiO}$ was varied. Presumably this is due to the surface of the SiO forming an $SiO_2$ crust in the presence of ozone, causing the flux to plummet. This is consistent with the results reported by Ardenghi *et al.*[36] for SiO used in conjunction with an oxygen plasma in the growth of silicon-doped $\beta$-$Ga_2O_3$ by plasma-assisted MBE (PAMBE). For this reason, we found SiO to be unsuitable as a controlled and reproducible doping source and moved on to evaluating $SiO_2$ for this purpose.

In calibrating the $SiO_2$ source, XRR measurements of amorphous $SiO_x$ on (0001) $Al_2O_3$ substrates revealed that a well-behaved and reproducible silicon flux emanates from it, despite changing the oxidant pressure from vacuum to $P_{O_3} = 2.5\times10^{-6}$ Torr. From the XRR spectra, the density of the amorphous $SiO_x$ films deposited by the $SiO_2$ source was about 2.1 g/cm$^3$ as tabulated for each sample in the supplementary material (Table S1). Having established that the $SiO_2$ source performed well when its temperature, $T_{SiO_2}$, was high, we moved on to evaluating its stability at lower $T_{SiO_2}$ relevant to doping $\beta$-$Ga_2O_3$ films. For these lower fluxes of the silicon-containing dopant species (mainly SiO from vapor pressure calculations[19]), Hall measurements were performed on silicon-doped homoepitaxial $\beta$-$Ga_2O_3$ films in both the oxidant-rich and adsorption-controlled regimes. The silicon flux is determined from the Hall measurements assuming full activation at room temperature, an assumption that may underestimate the silicon flux. On the other hand, mobile carriers due to the unintentional silicon contamination that often occurs at the interface between the $\beta$-$Ga_2O_3$ film and underlying substrate[14,25,43-45] could lead to the Hall measurements overestimating the silicon flux. Tables II and S1 (supplementary materials) list the growth conditions of the samples grown using an $SiO_2$ source. Table II lists the samples investigated by Hall measurements (Samples a – l) while Table S1 lists the samples measured by XRR (Samples o – v).

Figure 2 shows an Arrhenius plot of the silicon flux calculated from both the XRR and Hall data as a function of $T_{SiO_2}$. Clear Arrhenius behavior with an activation energy of about 5 eV is seen between the silicon flux and $1/T_{SiO_2}$. To assess our ability to control silicon doping over the $10^{16}$–$10^{19}$ $cm^{-3}$ range using the $SiO_2$ suboxide doping source, we attempted to grow two 7 µm thick films for SIMS analysis, one in the oxidant-rich regime and the other in the adsorption-controlled regime. Each film starts off with a 1 µm thick undoped $\beta$-$Ga_2O_3$ buffer layer followed by alternating 0.5 µm thick layers of silicon-doped and undoped $\beta$-$Ga_2O_3$ as shown by the schematics in Figs. 3(a) and 3(c) for the SIMS stacks grown in the oxidant-rich and adsorption-controlled regime, respectively. To keep the growth conditions similar yet explore the two growth regimes to see whether they affect the dopant incorporation or the doping profile in some way, both SIMS stacks were grown at the same $T_{sub}$ = 525 °C and same $P_{O_3}$ = $2.5\times10^{-6}$ Torr.

Accessing the oxidant-rich regime vs the adsorption-controlled regimes was achieved by using different $Ga_2O$ fluxes for these two SIMS stacks: $6\times10^{14}$ molecules $cm^{-2}$ $s^{-1}$ for the oxidant-rich regime and $1\times10^{15}$ molecules $cm^{-2}$ $s^{-1}$ for the adsorption-controlled regime (Fig. 1(a)). The growth rates in these two regimes for the growth conditions used differ (0.86 vs 1.3 μm/h), requiring different $T_{SiO_2}$ values. During the growth of each undoped layer, the temperature of the $SiO_2$ source is increased to provide the silicon flux targeted for the next silicon-doped layer.

Figures 3(b) and 3(d) are the resulting SIMS measurements depicting the concentration of elements of interest—silicon, iron, aluminum, and beryllium—with respect to the sputtered depth. The SIMS data show clear and well-defined steps, demonstrating controllable silicon-doping from $5\times10^{16}$ atoms $cm^{-3}$ to $1\times10^{19}$ atoms $cm^{-3}$. The steepness of the step edges of the silicon profile differs between the oxidant-rich and adsorption-controlled regimes, with the oxidant-rich regime consistently showing sharper steps. The underlying reason could be the increased rms roughness of the ~9 μm thick SIMS stack grown in the adsorption-controlled regime (53 nm) in comparison to the ~6.5 μm thick SIMS stack grown in the oxidant-rich regime (11 nm). We also note that when the silicon shutter is open, the silicon profile is also observed to be flatter for the oxidant-rich regime (Fig. 3(b)) whereas it has a slight upward slope (indicative of silicon riding the surface) for the adsorption-controlled regime (Fig. 3(d)).

Despite calibrating the silicon incorporation by XRR and Hall measurements, the silicon concentration measured by SIMS in Figs. 3(b) and 3(d) does not precisely match the incorporation targeted in our SIMS stack design in Figs. 3(a) and 3(c). At high silicon-doping ($T_{SiO_2} > 1187$ °C), the match is good, but at lower $T_{SiO_2}$, the concentration observed by SIMS is lower than the predicted silicon incorporation by up to a factor of 5. One explanation of this discrepancy is that the Hall data overestimate the silicon concentration at low doping because of the unintentional silicon contamination at the substrate interface[14,25,43-45] giving rise to a concentration of mobile carriers beyond that provided by the silicon flux during growth. To assess this hypothesis, additional SIMS measurements were conducted on two Hall samples (Samples k and e), one in each growth regime. If this hypothesis were true, then the concentration of silicon measured by SIMS should be less than the mobile carrier concentration measured on the same samples by the Hall effect. The results are shown in Fig. 2, where the vertical arrows point to the silicon fluxes measured by SIMS and Hall effect on the same sample. The data do not support the hypothesis. Rather, the data show that the silicon concentration measured by SIMS is as large or larger than the concentration measured by the Hall effect. The data show that at high doping, almost all of the silicon is electrically active, ~90% at $3\times10^{19}$ cm$^{-3}$ doping. At lower doping, however, the fraction of silicon dopants that produce mobile carriers is far lower, ~40% silicon activity at $2\times10^{18}$ cm$^{-3}$ doping. This could be from a background of compensating acceptor states in our films, but the concentration involved ~$10^{18}$ cm$^{-3}$ of acceptors or traps is so high that it is inconsistent with the relatively high mobilities seen in other films as we describe below.

The concentration of silicon incorporated into the $\beta$-$Ga_2O_3$ films as a function of $T_{SiO_2}$ in the SIMS stacks shown in Figs. 3(b) and 3(d) was used to calculate the silicon flux at doping-relevant temperatures and are also plotted in Fig. 2. From these SIMS values and the XRR values a more accurate fit to the Arrhenius behavior between the silicon flux and $1/T_{SiO_2}$ can be obtained. Unlike the Hall measurement, which only probes mobile carriers, the SIMS and XRR data measure all of the silicon incorporated. The resulting best fit has an activation energy of 5.3 eV/molecule. Although some scatter exists, particularly between Hall and SIMS values, the results in Fig. 2 show a linear trend that extends over 5 orders of magnitude, establishing that $SiO_2$ is a well-behaved doping source for the growth of $\beta$-$Ga_2O_3$ by suboxide MBE. We attribute the more stable behavior of the $SiO_2$ doping source over an SiO or silicon doping source to the fact that $SiO_2$ is fully oxidized and free of the active/passive oxidation issues[30-32] that plague SiO[36] and silicon[5,29] sources when used at the high oxidant pressures involved in the growth of $\beta$-$Ga_2O_3$ by MBE.

In addition to measuring the concentration of silicon, the concentrations of iron, aluminum, and beryllium were also measured by SIMS. This is because these elements are the major impurities in our $Ga_2O_3$ powder or they are the major constituents of the BeO and $Al_2O_3$ crucibles used to contain the $Ga_2O_3$ + gallium mixture and $SiO_2$ sources, respectively. Composition analysis on the 99.999% $Ga_2O_3$ powder reveals aluminum, boron, sodium, and iron to be the only elements present at above the ppm level. A broad screening of these elements and more by SIMS analysis (not shown) indicated that the major contaminants in our undoped $\beta$-$Ga_2O_3$ films grown by *S*-MBE are aluminum, silicon, and iron; all impurities that are not isoelectronic with gallium have concentrations below $10^{16}$ $cm^{-3}$.[14] The low concentration of beryllium in the SIMS in Figs. 3(b) and 3(d) demonstrate that BeO is a suitable crucible for the $Ga_2O_3$ + gallium mixture. Of concern, however, are the high iron levels seen in Figs. 3(b) and 3(d). The steps in iron contamination seen by SIMS track the temperature changes and shutter opening of the $SiO_2$ source. This indicates that the iron contamination is coming from either (1) impurities in the $SiO_2$ source material, with its 99.99% purity; (2) the 99.8% pure $Al_2O_3$ crucible used to contain the $SiO_2$; or (3) iron impurities from the iron-doped $\beta$-$Ga_2O_3$ substrate that ride the growth front[46] and are preferentially incorporated in the presence of silicon dopants.

When $SiO_2$ is heated to a temperature range of 1075–1490 °C, as is used in this work, the stable polymorph varies as is described in Fig. S1 of the supplementary material. Over this temperature range the silicon-containing species with the highest vapor pressure is the suboxide SiO, followed by $SiO_2$. According to our thermodynamic calculations (Fig. S1) the SiO makes up 61% of the silicon-containing species in the gas phase at 1075 °C and 99.3% at 1490 °C. Experimentally, the dominant silicon-containing species observed in the gas phase when $SiO_2$ is evaporated is SiO.[47-49] Further, the activation of energy of the measured vapor pressure of SiO, averaged from three studies (all with $f < 4\times10^{-3}$, where $f$ is a parameter characterizing the effusion cell and at this small magnitude is consistent with measurements of the equilibrium vapor pressure), is 5.29 ± 0.21 eV.[47-49] This agrees well with the activation energy of 5.3 eV from the fit in Fig. 2.

Having established that $SiO_2$ is a well-behaved doping source, 1 µm thick films of silicon-doped $\beta$-$Ga_2O_3$ are grown in both the oxidant-rich and adsorption-controlled regimes at ~1 µm/h. Figure 4(a) shows the mobility deduced from Hall measurements of silicon-doped $\beta$-$Ga_2O_3$ films grown by *S*-MBE (red stars) together with the best results from the literature of silicon-doped $\beta$-$Ga_2O_3$ films grown by other leading techniques.[3,5,24,27,50,51] From the mobility comparison it is evident that $\beta$-$Ga_2O_3$ grown by *S*-MBE with a $\beta$-$SiO_2$ doping source is on par with other leading techniques. Figure 4(b) distinguishes which films in Fig. 4(a) were grown in the oxidant-rich and adsorption-controlled regime by *S*-MBE. Interestingly, there is no clear difference between the mobility at room temperature of silicon-doped $\beta$-$Ga_2O_3$ grown by *S*-MBE in the oxidant-rich or adsorption-controlled regime.

The sample with the highest mobility at room temperature is Sample a, a 1 µm thick film grown by *S*-MBE with the $SiO_2$ doping source in the adsorption-controlled regime. At room temperature it has a mobility of 124 $cm^2\ V^{-1}\ s^{-1}$. The temperature-dependence of this mobility measured by Hall effect is shown in Fig. 5. The mobility peaks at 76 K at a mobility of 627 $cm^2 V^{-1} s^{-1}$. This value is significantly higher than all prior reports of $\beta$-$Ga_2O_3$ grown by MBE, i.e., higher than PAMBE[5] as well as MOCATAXY.[6] Using the temperature-dependent carrier density data for this sample as measured by Hall effect, an activation energy of 27 meV and a compensating acceptor density ($N_a$) of $4\times10^{15}$ $cm^{-3}$ was determined by fitting the temperature-dependent Hall carrier density using the charge neutrality equation.[52,53] This activation energy for silicon from our $SiO_2$ doping source is comparable to that of silicon in $\beta$-$Ga_2O_3$ grown by conventional MBE (with an elemental silicon doping source)[5,53] as well as MOCVD.[3,23,25]

The results of additional structural characterization measurements on this same high-mobility sample (Sample a) are shown in Fig. 6. The film is unremarkable when analyzed by XRD; single-phase epitaxial $\beta$-$Ga_2O_3$ is seen. Rocking curves of the 020 $\beta$-$Ga_2O_3$ peak reveal full widths at half maxima (FWHM) of 43 and 122 arcsec along two perpendicular in-plane directions. These values are on par with the rocking curves of the bare iron-doped $\beta$-$Ga_2O_3$ (010) substrates. AFM reveals an rms roughness of 1.8 nm for this 1 µm thick film grown at 0.86 µm/h. We thus demonstrate that using *S*-MBE for both the semiconductor and the dopant tackles some of the major challenges associated with the conventional MBE growth of $\beta$-$Ga_2O_3$.

Having demonstrated that *S*-MBE appears to be a viable method for the growth of $\beta$-$Ga_2O_3$ with good structural and electrical properties at record growth rates (for MBE), we return to the question of whether it can produce device-quality material. As an initial test, we prepared a simple MESFET on a 65 nm thick homoepitaxial $\beta$-$Ga_2O_3$ layer grown by *S*-MBE at a growth rate of ~1 µm/h with the structure shown in Fig. 7(a). For this test device, the SiO doping source was used to dope the epilayer. The MESFET results are shown in Figs. 7(b) and 7(c); although these are initial results, the ability of *S*-MBE to produce device-quality material is evident.

Device contacts were ohmic and the channel showed good modulation, with a peak transconductance ($G_m$) of 23.4 mS/mm, threshold voltage $V_{th}$ ~ –6 V, and $V_{off}$ = –13 V. The relatively low on/off ratio can be attributed to leakage through the parasitic channel formed by unintentional silicon contamination at the interface between the $\beta$-$Ga_2O_3$ film and underlying substrate.[14,25,43-45] These MESFETs perform comparably to devices with similar architectures,[40,54] indicating that the channel material grown by *S*-MBE is indeed suitable for device fabrication.

**Conclusions**

We have overcome the kinetic limitation to the slow growth rate of $\beta$-$Ga_2O_3$ by conventional MBE and demonstrate a growth rate of epitaxial $\beta$-$Ga_2O_3$ as high as 2.5 µm/h on (0001) $Al_2O_3$ substrates by *S*-MBE. By using a doping source in which silicon is in its most oxidized state, $SiO_2$, we avoid the oxidation issues of previously used silicon[4,5,29] and SiO dopant sources[36] when used at high oxidants pressures. This enables controllable and reproducible silicon-doping of $\beta$-$Ga_2O_3$ in the $5\times10^{16}$ $cm^{-3}$ to $10^{19}$ $cm^{-3}$ range for the high growth rates used. The doped films grown at ~1 µm/h exhibit mobilities at room temperature rivaling leading techniques and mobilities at low temperature that are the highest achieved to date by MBE, 627 $cm^2$ $V^{-1}$ $s^{-1}$ at 76 K. While these characteristics are still inferior to the electrical properties of $\beta$-$Ga_2O_3$ grown by MOCVD,[2-4,23-25] *S*-MBE is emerging as a viable technique for the growth of electronic-grade $\beta$-$Ga_2O_3$ at rates enabling the intensive investigation of thick vertical heterostructures.

**Acknowledgments**

K.A., C.A.G., N.A.P., J.S., J.P.M., D.J., H.G.X., D.A.M., M.O.T., H.P.N., and D.G.S. acknowledge support from the AFOSR/AFRL ACCESS Center of Excellence under Award No.

FA9550-18-1-0529. J.P.M. also acknowledges support from the National Science Foundation within a Graduate Research Fellowship under Grant No. DGE-1650441. P.V. and Y.A.B. acknowledge support from ASCENT, one of six centers in JUMP, a Semiconductor Research Corporation (SRC) program sponsored by DARPA. F.V.E.H. acknowledges support from the Alexander von Humboldt Foundation in the form of a Feodor Lynen fellowship. F.V.E.H. also acknowledges support from the National Science Foundation (NSF) [Platform for the Accelerated Realization, Analysis and Discovery of Interface Materials (PARADIM)] under Cooperative Agreement No. DMR-1539918. M.D.W., D.A.M., and D.G.S. acknowledge support from the NSF under DMR-2122147. M.D.W. also acknowledges NSF HRD-1924204 and ONR Award N00014-21-1-2823. This work made use of the Cornell Center for Materials Research (CCMR) Shared Facilities, which are supported through the NSF MRSEC Program (Grant No. DMR-1719875). Substrate preparation was performed, in part, at the Cornell NanoScale Facility, a member of the National Nanotechnology Coordinated Infrastructure (NNCI), which is supported by the NSF (Grant No. NNCI-2025233).

The authors K.A., F.V.E.H., S.-L.S., P.V., Z.-K.L., and D.G.S. have been granted U.S. Patent No. 11,462,402 (4 October 2022) with the title “Suboxide Molecular-Beam Epitaxy and Related Structures.”

**Data Availability**

The data supporting the findings of this study are available within the paper. Additional data related to the growth and structural characterization are available at https://doi.org/10.34863/???. Any additional data connected to the study are available from the corresponding author upon reasonable request.

**Tables**

Table I. Comparison of the highest reported electrical mobilities of $\beta$-$Ga_2O_3$ films grown at relatively high rate by MBE vs MOCVD.

| | Mobility at Room Temperature ($cm^2$ $V^{-1}$ $s^{-1}$) | Peak Mobility ($cm^2$ $V^{-1}$ $s^{-1}$) | Temperature of Peak Mobility (K) | Growth Rate (µm/h) | Ranged of Controlled Doping ($cm^{-3}$) | Reference |
|---|---|---|---|---|---|---|
| MBE[a] | 129 | 390 | 97 | 0.09 | $10^{17}$ - $10^{20}$ | 5 |
| MBE[b] | 136 | 168 | 165 | 0.3 | $4\times10^{16}$ - $2\times10^{19}$ | 6 |
| MOCVD | 190 | 3425 | 53 | 2.9 | $2\times10^{16}$ - $4\times10^{19}$ | 3 |

[a] Conventional plasma-assisted MBE

[b] Metal-oxide catalyzed epitaxy (MOCATAXY)

Table II. Table of growth parameters and electrical characteristics of the $\beta$-$Ga_2O_3$ films measured by the Hall effect and SIMS.[a]

| Name | Mobility ($cm^2$ $V^{-1}$ $s^{-1}$) | Carrier Density ($cm^{-3}$) | $SiO_2$ Source Temperature (°C) | Thickness (nm) | $Ga_2O$ Flux (molecules $cm^{-2}$ $s^{-1}$) |
|---|---|---|---|---|---|
| Sample a | 124 | $2.7\times10^{17}$ | 1286 | 1000 | $1\times10^{15}$ |
| Sample b | 119 | $2.9\times10^{17}$ | 1134 | 1000 | $6\times10^{14}$ |
| Sample c | 111 | $3.5\times10^{17}$ | 1115 | 1000 | $6\times10^{14}$ |
| Sample d | 129 | $3.5\times10^{17}$ | 1119 | 1000 | $1\times10^{15}$ |
| Sample e | 104 | $9.8\times10^{17}$ | 1187 | 1000 | $6\times10^{14}$ |
| Sample f | 86 | $1.3\times10^{18}$ | 1225 | 1000 | $1\times10^{15}$ |
| Sample g | 98 | $1.5\times10^{18}$ | 1171 | 1000 | $6\times10^{14}$ |
| Sample h | 91 | $2.4\times10^{18}$ | 1176 | 1000 | $1\times10^{15}$ |
| Sample i | 75 | $5.5\times10^{18}$ | 1262 | 1000 | $1\times10^{15}$ |
| Sample j | 79 | $1.1\times10^{19}$ | 1286 | 1000 | $1\times10^{15}$ |
| Sample k | 62 | $3.0\times10^{19}$ | 1319 | 1000 | $1\times10^{15}$ |
| Sample l | 68 | $4.4\times10^{19}$ | 1350 | 1000 | $1\times10^{15}$ |
| Sample m | - | - | Figure 3(a) | 6500 | $6\times10^{14}$ |
| Sample n | - | - | Figure 3(c) | 9000 | $1\times10^{15}$ |

[a] All samples were grown at $T_{sub}$ = 525 °C in $P_{O_3}$ = $2.5\times10^{-6}$ Torr on iron-doped (010) $\beta$-$Ga_2O_3$ substrates and doped with an $SiO_2$ source.

**Figures and Figure Captions**

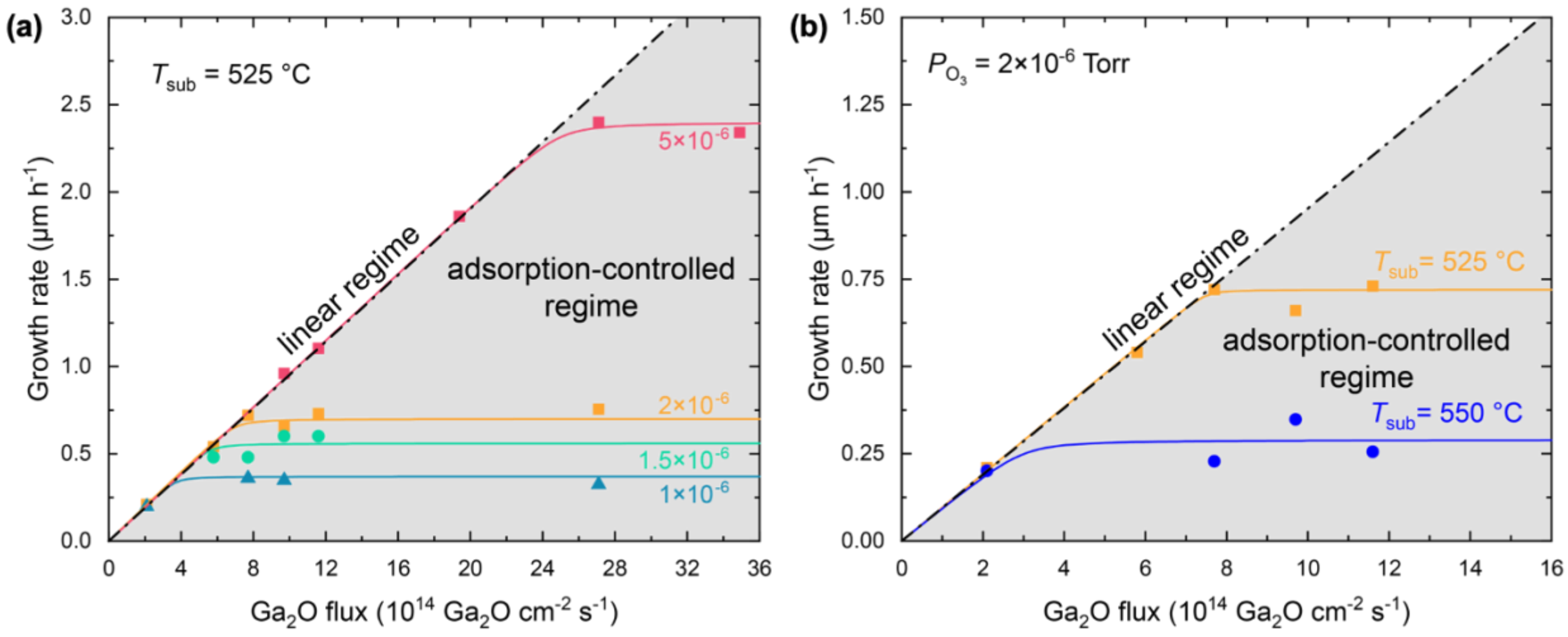

FIG. 1: Growth rate of $\beta$-$Ga_2O_3$ on (0001) $Al_2O_3$ substrates as a function of $Ga_2O$ flux. (a) Growth rate at a constant substrate temperature ($T_{sub}$ = 525 °C) at four different background pressures (in Torr) of distilled ozone showing the oxidant-rich (linear) regime and the $Ga_2O$-rich (adsorption-controlled) regime. (b) The effect of substrate temperature on the growth rate. The smooth curves indicate a fit to the experimental growth rates by the kinetic model described in Ref. 33.

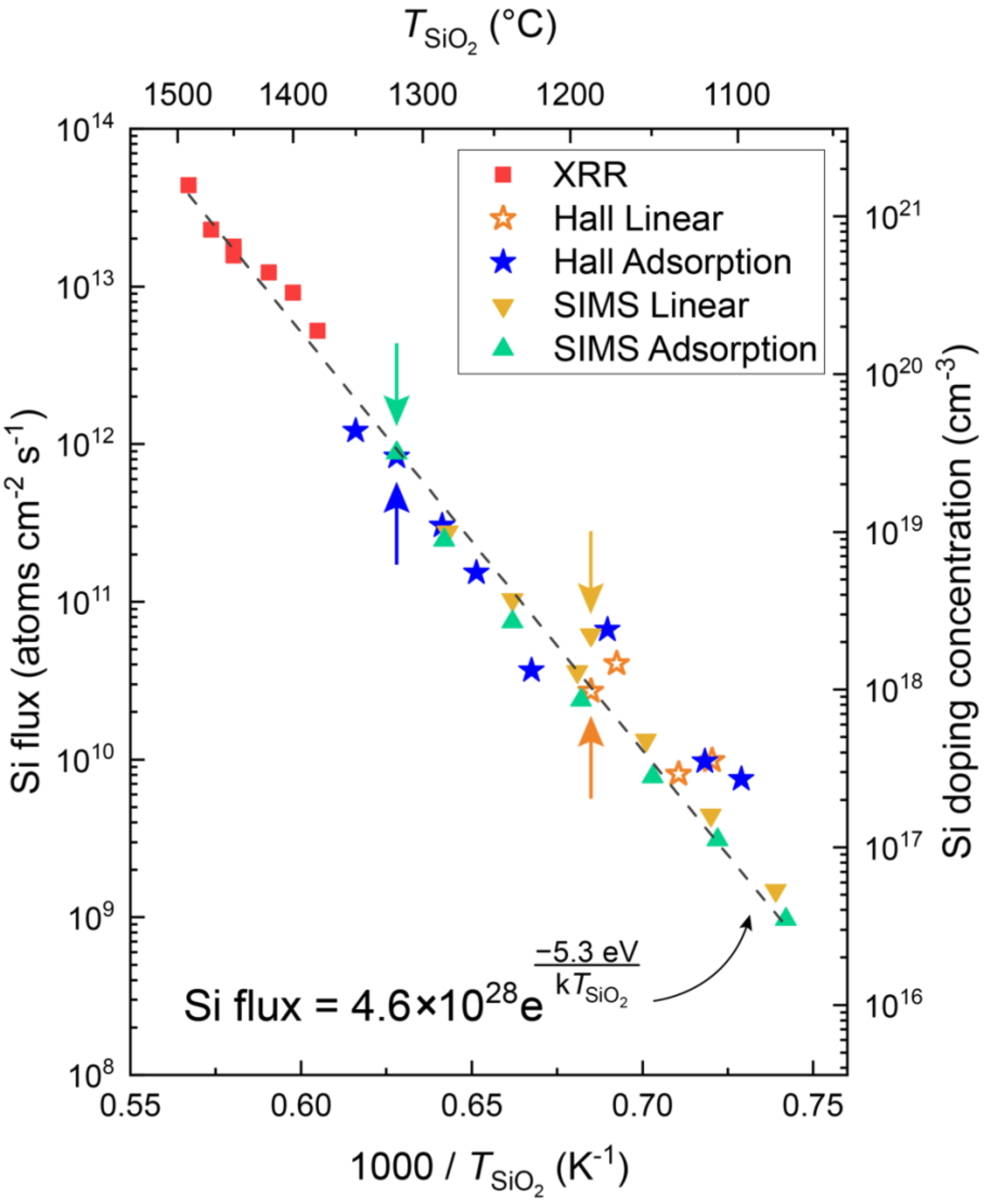

FIG. 2: Silicon flux incorporated into $\beta$-$Ga_2O_3$ films as measured by XRR (red squares), SIMS (triangles), and Hall effect (stars) as a function of the temperature of the $SiO_2$ doping source ($T_{SiO_2}$). The linear fit shown is calculated based on the XRR and SIMS data points, where $T_{SiO_2}$ is in Kelvin. Two of the samples on which Hall effect was measured were also measured by SIMS. For those two samples the calculated silicon flux values are indicated by the vertical arrows, showing the fraction of silicon that is electrically active is lower when the $\beta$-$Ga_2O_3$ is less highly doped. The vertical axis on the right side of the figure gives the silicon doping concentration that the silicon flux would produce in a film of $\beta$-$Ga_2O_3$ grown at 1 μm/h.

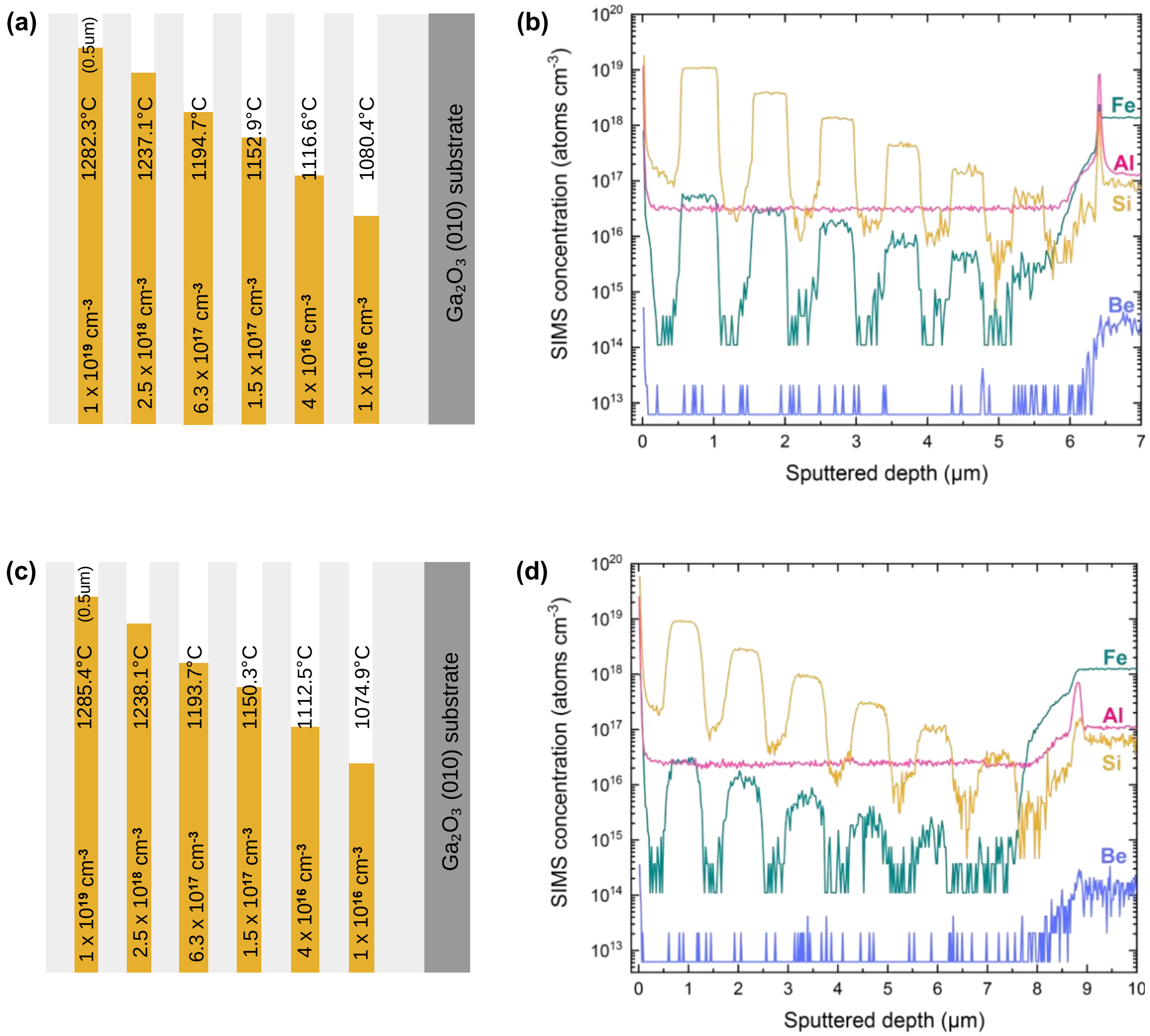

FIG. 3: SIMS profiles showing the concentrations of silicon, iron, aluminum, and beryllium as a function of depth of silicon-doped $\beta$-$Ga_2O_3$ films grown at $T_{sub}$ = 525 °C and a background oxidant pressure of $2.5\times10^{-6}$ Torr of distilled ozone. (a) Schematic of the targeted doping profile, including $T_{SiO_2}$ for each layer and (b) the measured SIMS profile of silicon-doped $\beta$-$Ga_2O_3$ films grown in the oxidant-rich (linear) regime at a growth rate of 0.86 μm/h at a $Ga_2O$ flux of $6\times10^{14}$ molecules $cm^{-2}$ $s^{-1}$. (c) Schematic of the targeted doping profile including $T_{SiO_2}$ for each layer and (d) the measured SIMS profile of silicon-doped $\beta$-$Ga_2O_3$ films grown in the $Ga_2O$-rich (adsorption-controlled) regime at a growth rate of 1.3 μm/h at a $Ga_2O$ flux of $1\times10^{15}$ molecules $cm^{-2}$ $s^{-1}$.

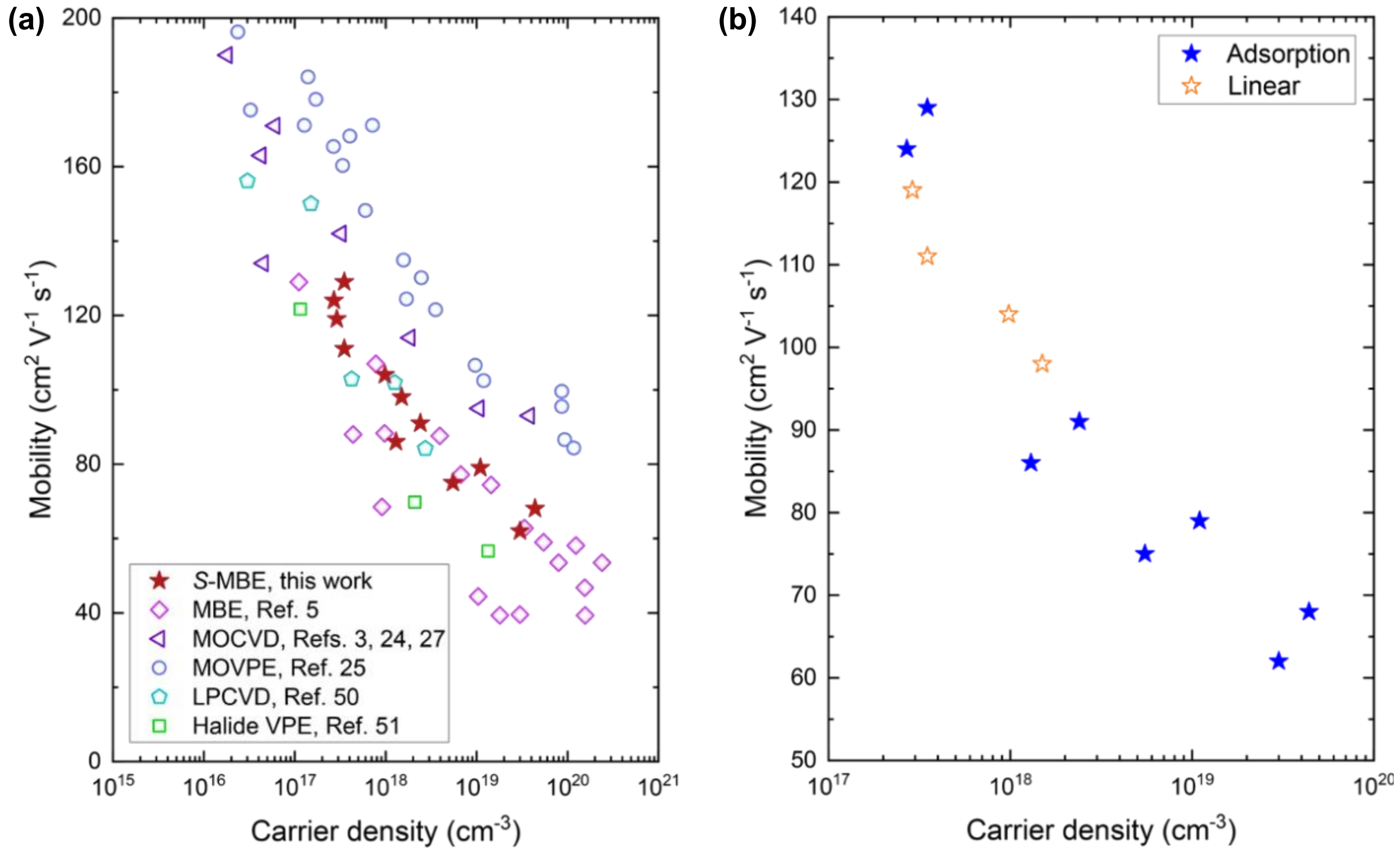

FIG. 4: (a) Comparison of the electron mobility as a function of electron density measured at room temperature by Hall effect on the silicon-doped $\beta$-$Ga_2O_3$ films grown in this study grown at ~1 µm/h by *S*-MBE to leading reports of silicon-doped $\beta$-$Ga_2O_3$ films by other techniques from the literature. The *S*-MBE results are indicated by the red stars. (b) A close-up of the *S*-MBE results (the results indicated by the red stars in (a)) showing which films were grown in the oxidant-rich regime (hollow stars) and which were grown in the adsorption-controlled regime (solid stars).

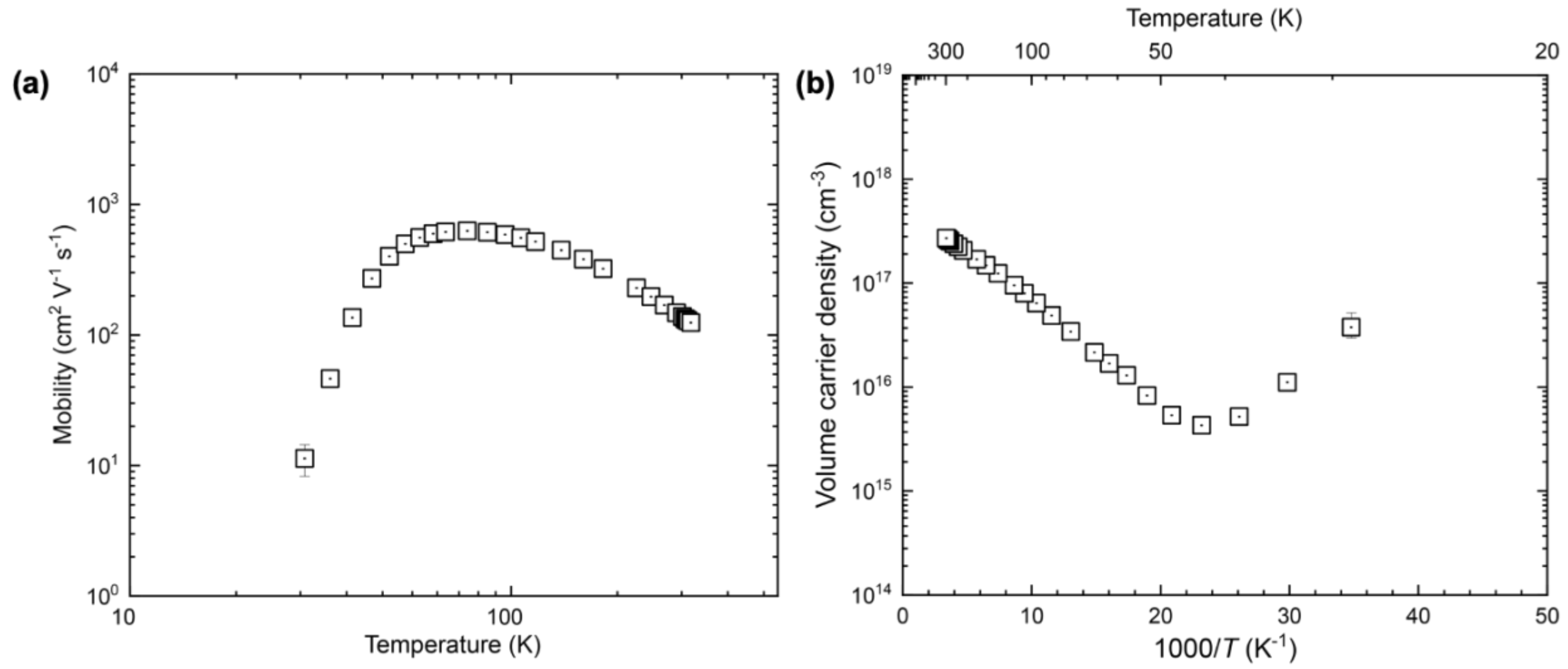

FIG. 5: Temperature-dependent Hall effect measurements made on a 1 μm thick film (Sample a) grown by *S*-MBE at a growth rate of ~1 μm/h with a mobile carrier concentration of $2.7\times10^{17}$ cm$^{-3}$ and a room-temperature mobility of 124 cm$^2$ V$^{-1}$ s$^{-1}$. The mobility of the charge carriers (electrons) (a) and their density (b) are shown as a function of temperature. The mobility peaks at 76 K at a mobility of 627 cm$^2$ V$^{-1}$ s$^{-1}$.

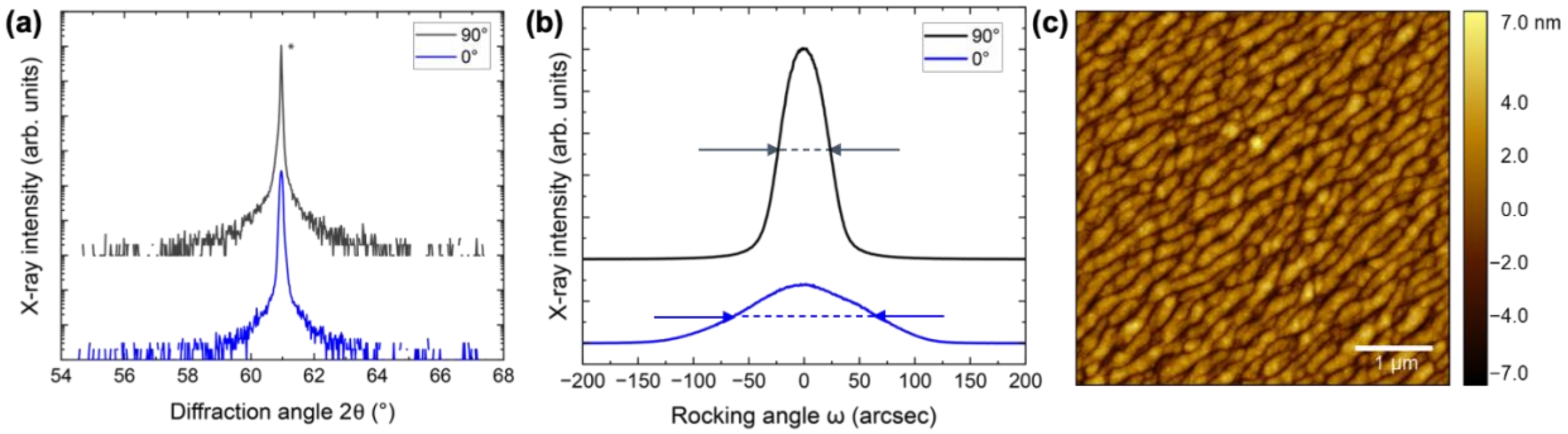

FIG. 6: Structural characterization of the same film shown in Fig. 5 by (a) $\theta$-2$\theta$ XRD scan in the vicinity of the 020 $\beta$-$Ga_2O_3$ peak where the scans along $\phi = 0°$ and $\phi = 90°$ are offset for clarity. (b) rocking curve of the 020 $\beta$-$Ga_2O_3$ peak showing a FWHM of 122 arcsec (blue) and 43 arcsec (grey) along $\phi = 0°$ and $\phi = 90°$, respectfully; the scans are offset for clarity. (c) AFM scan of the same film revealing an rms roughness of 1.9 nm.

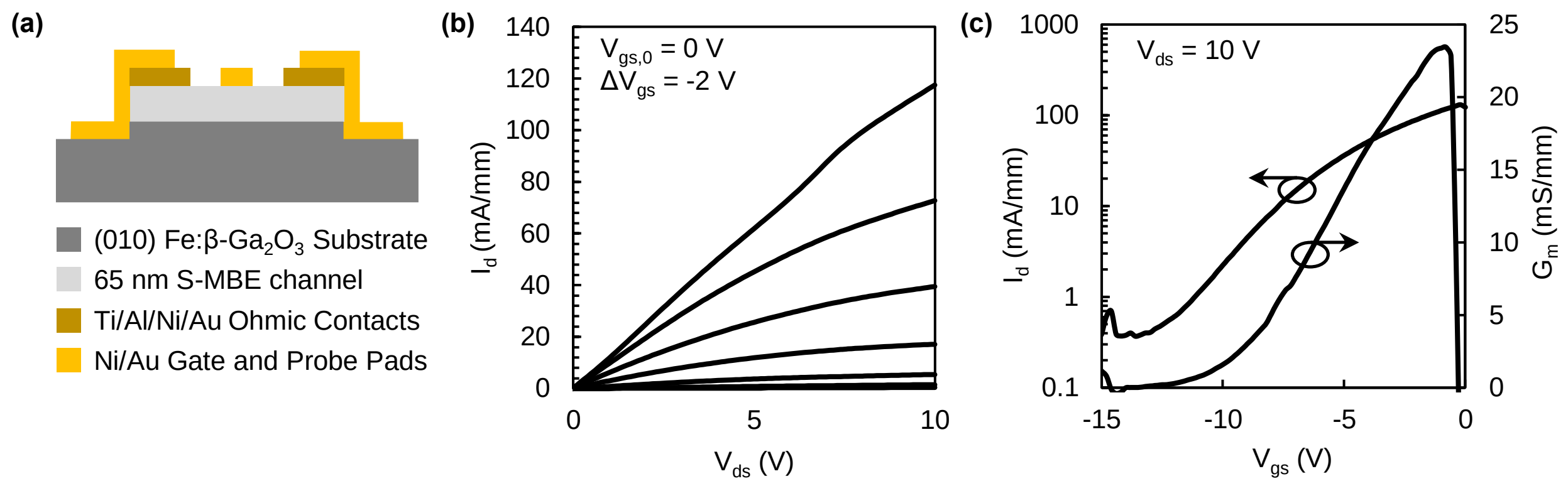

FIG. 7: a) Schematic of the annealed-ohmic MESFET fabricated using a silicon-doped $\beta$-$Ga_2O_3$ film grown at ~1 μm/h by *S*-MBE with $N_d = 10^{18}$ cm$^{-3}$. b) Output curves at $V_{gs}$ = 0, −2, –4, –6, –8, and −10 V and c) transfer curve and transconductance of the MESFET.

Supplementary Material

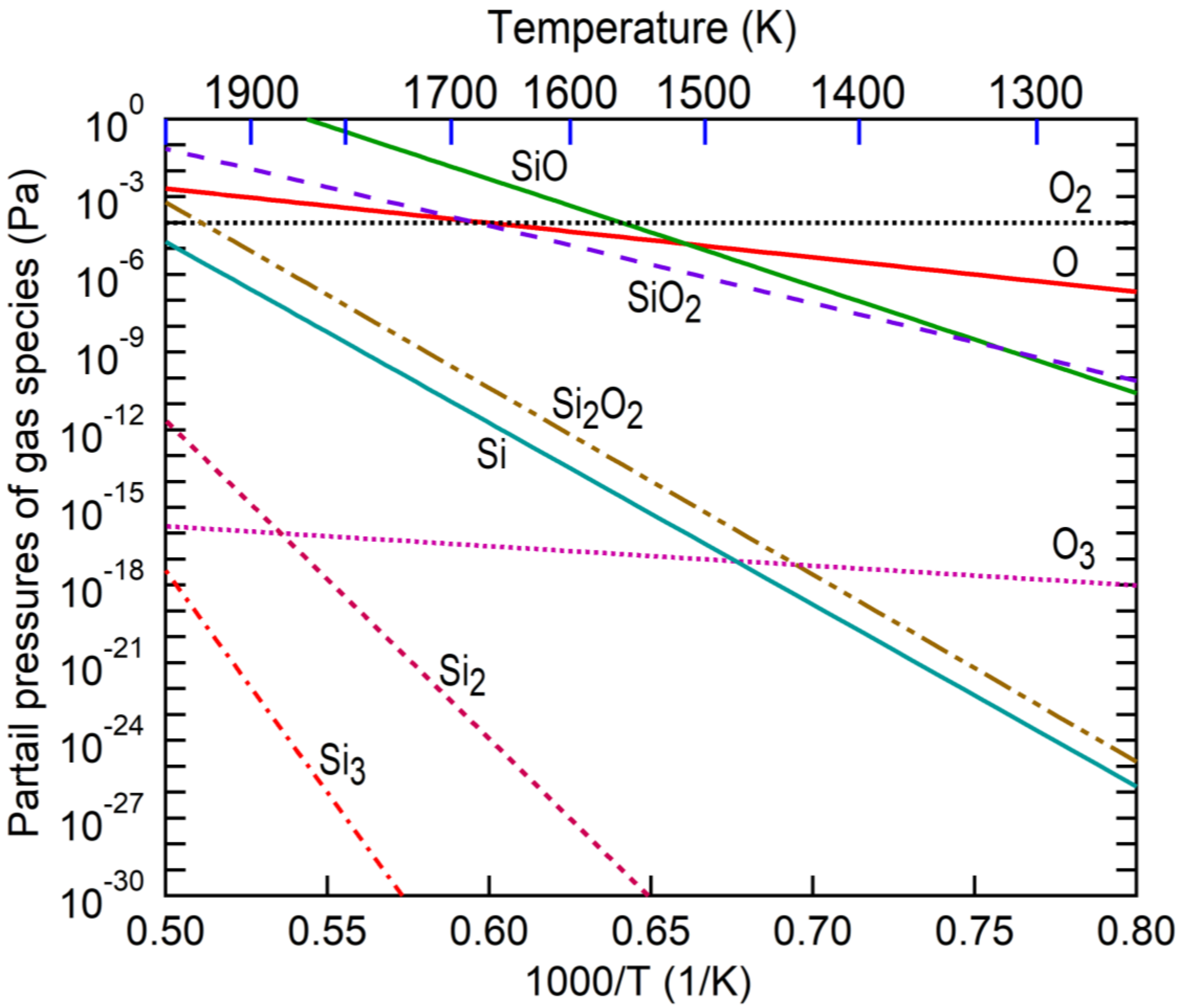

Figure S1. Calculated partial pressure of the species in the gas phase over $SiO_2$_tridymite as the source material as a function of temperature at a fixed oxygen partial pressure of $7.6\times10^{-7}$ Torr ($10^{-4}$ Pa). Note that (i) the stable $SiO_2$ phases are quartz for temperature $T < 847$ K, *b*-quartz for $847 < T < 1143$ K, tridymite_S3 (one of the polymorphs of tridymite) for $1143 < T < 1746$ K, cristobalite for $1746 < T < 1996$, and liquid $SiO_2$ for $T > 1996$ K; (ii) the partial pressures of these gas species are similar by using different $SiO_2$ solids as source materials; and (iii) other details of thermodynamic calculations are given in Ref. 19.

Table S1. Table of growth parameters and growth rates by XRR of amorphous $SiO_2$ calibration films grown on (0001) $Al_2O_3$ substrates.

| Name | $SiO_2$ Source Temperature (°C) | Substrate Temperature (°C) | Thickness (nm) | Growth Time (min) | Growth Rate (µm/h) | Si Flux (atom cm$^{-2}$ s$^{-1}$) | Density (g/cm$^3$) | $P_{O_3}$ (Torr) |
|---|---|---|---|---|---|---|---|---|
| Sample o | 1400 | 20 | 31 | 300 | 0.006 | $9.1\times10^{12}$ | $2.5^{+0.16}_{-0.04}$ | $2.5\times10^{-6}$ |
| Sample p | 1450 | 20 | 42 | 180 | 0.014 | $1.8\times10^{13}$ | $2.1^{+0.16}_{-0.05}$ | $2.5\times10^{-6}$ |
| Sample q | 1450 | 20 | 39 | 171 | 0.014 | $1.6\times10^{13}$ | $2.0^{+0.29}_{-0.05}$ | $2.5\times10^{-6}$ |
| Sample r | 1450 | 20 | 42 | 180 | 0.014 | $1.6\times10^{13}$ | $1.9^{+0.08}_{-0.05}$ | None |
| Sample s | 1470 | 20 | 57 | 180 | 0.019 | $2.3\times10^{13}$ | $2.0^{+0.22}_{-0.07}$ | None |
| Sample t | 1420 | 20 | 27 | 180 | 0.027 | $1.2\times10^{13}$ | $1.9^{+0.08}_{-0.05}$ | None |
| Sample u | 1490 | 20 | 34 | 60 | 0.039 | $4.4\times10^{13}$ | $2.3^{+0.07}_{-0.05}$ | None |
| Sample v | 1380 | 20 | 23 | 360 | 0.004 | $5.2\times10^{12}$ | $2.3^{+0.09}_{-0.05}$ | None |